\documentclass[reprint,superscriptaddress,amsmath,amssymb,aps, prb,floatfix]{revtex4-2}
\usepackage{amsmath}
\usepackage{amssymb}
\usepackage{hyperref}
\usepackage{amsfonts}
\usepackage{amsmath}
\usepackage{amssymb}
\usepackage{graphicx}
\usepackage{color}
\usepackage{tensor}
\usepackage{upgreek}
\usepackage{accents}
\usepackage[percent]{overpic}
\usepackage{times}
\usepackage{subfigure}
\usepackage{latexsym}
\usepackage{bm}
\usepackage{anyfontsize}
\usepackage{hyperref}
\usepackage{multirow}
\usepackage{booktabs}
\usepackage{makecell}

\begin{document}

\title{ $\pmb{d}$-vector precession induced pumping in topological $p$-wave superconductors}
\author{Jun-Jie Fu}
\affiliation{National Laboratory of Solid State Microstructures, Department of Physics, Nanjing University, Nanjing 210093, China}
\author{Jin An}
\email{anjin@nju.edu.cn}
\affiliation{National Laboratory of Solid State Microstructures, Department of Physics, Nanjing University, Nanjing 210093, China}
\affiliation{Collaborative Innovation Center of Advanced Microstructures, Nanjing University, Nanjing 210093, China}
\begin{abstract}
    Time-reversal invariant $p$-wave superconductors (SCs) are characterized by their $\bm{d}$-vectors, whose orientations could be manipulated by a tiny magnetic field. We study in this paper the adiabatic pumping process induced by periodically rotating $\bm{d}$-vector in a topological $p$-wave SC, which is coupled to two normal leads. If $\bm{d}$-vector rotates nearly within a plane, the pumped spin $2S_z/\hbar$ over one cycle is nearly quantized at $2$ without net charge pumping. When the pumping lead is fully spin-polarized, both the pumped charge $Q/e$ and spin $2S_z/\hbar$ would peak nearly at $1$. When a mixing $s$-wave pairing component is taken into account, a topological phase transition can be driven by modulating the ratio between the pairing components. We found a sharp resonance phenomenon near the phase transition when the $p$-wave $\bm{d}$-vector is adiabatically rotating, which may help experimentally distinguish the topological SCs from trivial ones.  
\end{abstract}

\maketitle
\section{Introduction}
Majorana zero modes (MZMs), existing in topological superconductors (SCs)\cite{PhysRevB.44.9667,T.M.Rice_1995,PhysRevB.61.10267,Kitaev_2001,PhysRevB.78.195125,PhysRevLett.100.096407,PhysRevLett.102.187001,PhysRevLett.105.097001,PhysRevLett.111.056402,PhysRevB.90.045118} and exhibiting non-Abelian exchange statistics\cite{PhysRevLett.86.268,RevModPhys.80.1083,PhysRevLett.104.040502,Alicea_2012,PhysRevX.4.021018,PhysRevX.6.031016,10.21468/SciPostPhysLectNotes.15,PhysRevLett.129.227002,PhysRevB.105.054507,PhysRevLett.131.176601,PRXQuantum.5.010323}, are expected to have potential applications in topological quantum computations. How to detect or confirm them experimentally has been a challenging problem in recent years\cite{PhysRevLett.103.237001,PhysRevB.82.180516,PhysRevLett.110.126406,doi:10.1126/science.1222360,PhysRevLett.119.136803,PhysRevResearch.2.013377,Nat.Phy.19.165,PhysRevLett.98.237002,PhysRevLett.101.120403,PhysRevLett.111.036802,PhysRevB.91.081405,PhysRevB.105.205430,PhysRevB.111.L121401,PhysRevB.79.161408,NatPhys.8.11,PhysRevLett.112.037001,PhysRevLett.114.166406,PhysRevLett.116.257003,science.358.6364}. Rashba spin-orbit nanowires, in proximity to an $s$-wave SC and subjected to a time-reversal-breaking Zeeman field\cite{PhysRevLett.105.077001,PhysRevLett.105.177002,doi:10.1126/science.1222360,PhysRevLett.108.096802,PhysRevResearch.2.013377,PhysRevB.107.245423,PhysRevB.107.035427,Nat.Phy.19.165,PhysRevLett.133.266605} or further in proximity to an altermagnet\cite{PhysRevLett.133.106601,PhysRevB.111.L121401}, can host effective $p$-wave pairing and provide a promising experimental platform to generate MZMs. Although the total Andreev reflection and the resonant zero-bias conductance peak are essential features for the existence of MZMs\cite{PhysRevLett.103.237001,PhysRevB.82.180516,PhysRevLett.110.126406,doi:10.1126/science.1222360,PhysRevLett.119.136803,PhysRevResearch.2.013377}, they do not correspond uniquely to the latter\cite{PhysRevB.96.075161,PhysRevB.97.165302,PhysRevB.98.155314,PhysRevLett.123.107703,NatPhys.17.4}. Non-stationary transports like periodic pumping give new approaches to detect MZMs\cite{PhysRevLett.111.116402,PhysRevB.88.140508,PhysRevB.89.045307,PhysRevB.99.085435,PhysRevB.103.195407}. Quantized charge pumping\cite{PhysRevB.27.6083,nphys913,PhysRevB.82.195440,PhysRevB.103.205410} or quantized spin pumping\cite{PhysRevB.82.161303,PhysRevLett.130.237002} have been found to be generated by precessing a Zeeman field. On the other hand, as a characteristic feature of a $p$-wave SC, the $\boldsymbol{d}$-vector is relatively locked in the effective $p$-wave nanowires. How the precession of this degree of freedom affects the pumping in a topological SC, especially in a time-reversal invariant SC has rarely been studied. Since the orientation of $\boldsymbol{d}$ is relevant to the intrinsic phases of the MZMs, the precession induced pumping should be capable of capturing new features of the MZMs. 

In this paper we focus on a one dimensional (1D) time-reversal symmetric $p$-wave SC coupled to two metallic chains, containing a Majorana Kramers pair at each interface. By rotating adiabatically the $p$-wave $\bm{d}$-vector, which can be manipulated by a tiny magnetic field, we study the charge and spin pumping in one of the leads. When the rotating $\bm{d}$-vector is nearly within a plane, over one cycle we found that nearly one $\downarrow$-hole are pumped in while simultaneously nearly one $\uparrow$-hole are pumped out. If the pumping lead is fully spin-polarized, both the pumped charge $Q/e$ and pumped spin $2S_z/\hbar$ would peak nearly at $1$. If a mixing $s+p$-wave pairing SC is considered instead, we found a sharp resonance phenomenon near the topological phase transition when adiabatically rotating the $p$-wave $\bm{d}$-vector. These phenomena would provide the smoking-gun signature of the existence of Kramers pair of MZMs, and may help determine experimentally the direction of the $p$-wave $\bm{d}$-vector in a time-reversal invariant SC. Furthermore, if the $p$-wave SC is fully spin polarized containing only one MZM at each end, a distinguished pumping phenomenon induced by rotating the magnetic field is also revealed. 

\begin{figure}[ht]
  \begin{center}
	\includegraphics[width=8.5cm,height=6.92cm]{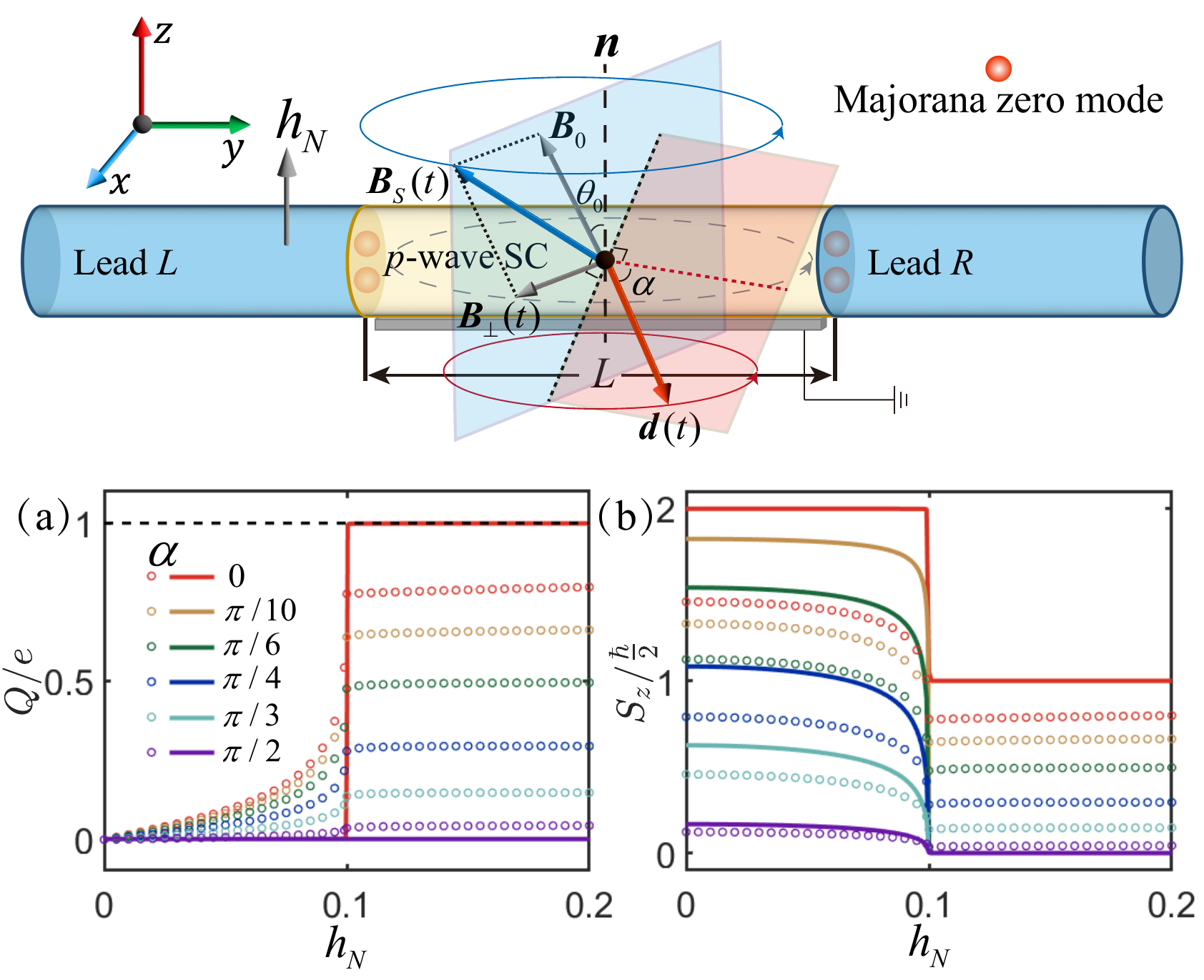}	
  \end{center}
  \vspace{-0.4cm}
	\caption{ Device designed to explore the periodic pumping in a 1D $p$-wave SC coupled by two normal leads, where a tiny slowly-varying magnetic field $\bm{B}_S(t)$ (consisting of both a rotating $\bm{B}_{\bot}(t)$ and constant $\bm{B}_0$ components) induces slow variation of the $p$-wave $\bm{d}$-vector, which is always perpendicular to the former. Pumped (a) charge and (b) spin in Lead $L$ in one cycle as functions of Zeeman field $h_N$ for different initial $\bm{d}$-vector's orientation $\alpha$, with $L=1000$ (300) for the solid lines (open circles). Parameters: $T=0$, $t_{NS}=-0.6$, $\mu=-1.9$, $\Delta_p=0.02$, $\eta=0.5$, $\theta_0=\pi/3$, and $\omega \to 0$. } 
  \label{fig1}
\end{figure}

Our pumping model system is schematically shown in Fig. \ref{fig1}, where two normal leads are coupled to a 1D $p$-wave SC, whose Hamiltonian is given by:
\begin{equation}
    H=\sum_{k,\sigma}[\xi_k c_{k\sigma}^{\dag}c_{k\sigma}
    +\Delta_p( i\sin k\ e^{i\delta_\sigma \alpha} c_{k\sigma}^{\dag}c_{-k\sigma}^{\dag}+\text{h.c.})],
\label{eq1}
\end{equation}where $\delta_{\uparrow/\downarrow}=\pm1$, $\xi_k=-2\cos k-\mu$ is the normal dispersion with $\mu$ the chemical potential, and $\Delta_p$ is the nearest-neighbor $p$-wave pairing potential. The pairing matrix can be expressed as $(\bm{d}_k\cdot \bm{\sigma})i\sigma_y$, with the $\bm{d}$-vector chosen within the $xy$-plane, characterized by phase $\alpha$: $\bm{d}_k=\Delta_p\sin k(\sin \alpha,\cos \alpha, 0)$. At each interface a Kramers pair of MZMs occurs. Both leads are described by $\mathcal{H}_N=\xi_k\tau_z$ with $\boldsymbol{\tau}$ the particle-hole Pauli matrices, and a Zeeman field $\bm{h}_N$ (assumed always to be along $\bm{z}$) is also introduced in Lead $L$. The interface hopping integral is $t_{NS}$.

This paper is organized as follows. In Sec. \ref{sec2}, we give the numerical results of charge and spin pumping at zero temperature and the analytic results of reflection amplitudes related to the orientation of $\bm{d}$-vector. In Sec. \ref{sec3}, we discuss the impacts of temperature and MZM-induced interference effects. In Sec. \ref{sec4}, we discuss the pumping in a mixed $s+p$-wave SC and focus on behaviors near the topological phase transition. In Sec. \ref{sec5}, we give the numerical results of pumping in a fully spin-polarized $p$-wave SC. In Sec. \ref{sec6}, we make a further discussion on a more realistic effective $p$-wave model.
\section{Pumping in a $p$-wave SC}
\label{sec2}
We further introduce a tiny slowly-varying magnetic field $\bm{B}_S(t)$ (assumed to be along $\bm{z}$ at $t=-\infty$, consistent with Eq.(\ref{eq1})) in the SC, which consists of a constant component $\bm{B_0}=B_0(\sin \theta_0,0, \cos \theta_0)$ and a periodically rotating one $\bm{B_{\bot}}(t) = B_{\bot}(\cos \omega t,\sin \omega t,0)$, with $B_0/B_{\bot}$ fixed to be $\eta$ and $\omega \to 0$. In the adiabatic limit, the spin polarization of paired electrons is expected to instantly follow the orientation of the tiny magnetic field, hence the pairing term in Eq.(\ref{eq1}) at any moment is varied to be $i\Delta_p\sum_k \sin k (e^{i\alpha}c_{k\Uparrow}^{\dag}c_{-k\Uparrow}^{\dag}+e^{-i\alpha}c_{k\Downarrow}^{\dag}c_{-k\Downarrow}^{\dag})$, where $\Uparrow$ ($\Downarrow$) denotes the spin orientation along (against) the field. During adiabatic variation of $\bm{B}_S(t)$, in spherical coordinates given by $ (B_S,\theta_B,\phi_B)$, since $(c_{\Uparrow}^{\dag},c_{\Downarrow}^{\dag})=(c_{\uparrow}^{\dag},c_{\downarrow}^{\dag}) U(t)$, where $U(t)=U_z(\phi_B)U_y(\theta_B)$ with $U_{\bm{n}}(\theta)=\exp (-i\frac{\theta}{2} \bm{\sigma} \cdot \bm{n})$ being the spin rotation by $\theta$ around $\bm{n}$, at each instant the pairing matrix $\Delta$ becomes: $U(t)\Delta U^T(t)$, resulting in $\bm{d}$-vector being varied to be $\bm{d}_k=\Delta_p \sin k\hat{\bm{d}}(t)$. Here $\hat{\bm{d}}(t)=\hat{\bm{e}}\times\hat{\bm{B}}_S\sin \alpha+\hat{\bm{e}}\cos\alpha$, where $\hat{\bm{e}}=\hat{\bm{z}}\times \hat{\bm{B}}_S/|\hat{\bm{z}}\times \hat{\bm{B}}_S|$. Within one cycle, the area swept by $\bm{d}$-vector is a cone, which becomes a disk in $xy$-plane when $\alpha=0$.
We focus on the adiabatic pumping in Lead $L$. The pumped charge $Q$ and spin $S_z$ over one cycle can be obtained by\cite{PhysRevB.58.R10135,PhysRevB.65.235318,WorldScientific}:
\begin{equation}
    \begin{aligned}
        &Q = \int dE(-\frac{\partial f}{\partial E}
)\mathcal{Q}(E),\\
        &S_z = \int dE(-\frac{\partial f}{\partial E}
)\mathcal{S}_z(E),\\
       &\mathcal{Q}(E)=i\frac{e}{2\pi}\int_{0}^{2\pi/\omega}dt\sum_{\sigma,\sigma '}(R^{\sigma \sigma '}+T^{\sigma \sigma '}),\\
       &\mathcal{S}_z(E)=i\frac{\hbar}{4\pi}\int_{0}^{2\pi/\omega}dt\sum_{\sigma,\sigma '}\delta_{\sigma} (R^{\sigma \sigma '}+T^{\sigma \sigma '}),
    \end{aligned}
    \label{eq3}
\end{equation}
where $R^{\sigma \sigma '}=(r^{\sigma\sigma'}_{ee})^*\partial_t r^{\sigma\sigma'}_{ee}-(r^{\sigma\sigma'}_{he})^*\partial_t r^{\sigma\sigma'}_{he}$ ($T^{\sigma \sigma '}=(t'^{\sigma\sigma'}_{ee})^*\partial_t t'^{\sigma\sigma'}_{ee}-(t'^{\sigma\sigma'}_{he})^*\partial_t t'^{\sigma\sigma'}_{he}$) denotes the current contribution by reflection from (transmission through) left interface, with $r^{\sigma \sigma'}_{ee}$, $t'^{\sigma \sigma'}_{ee}$ ($r^{\sigma \sigma'}_{he}$, $t'^{\sigma \sigma'}_{he}$) the spin-dependent normal (Andreev) reflection and transmission amplitudes. Due to the time partial-derivatives, the integrand would be proportional to $\omega$, and the integrals over $t$ can then be expressed as an loop integrals over a geometrical parameter space, independent of $\omega$.

First, we focus on the case of $L\gg l_M$, with $l_M$ the attenuation length of the MZMs, where the transmissions tend to vanish and the pumping is nearly totally contributed by reflections from the left interface. The numerical results of pumping at $T=0$ in one cycle are shown by the solid lines in Fig. \ref{fig1}. When $h_N<h_c$, with $h_c = 2+\mu$ the critical normal Zeeman field, there is no charge pumping but the spin pumping is finite and shows $\alpha$-dependent feature, being quantized to be $\hbar$ at $\alpha=0$. While when Lead $L$ is fully spin-polarized, namely, $h_N>h_c$, the pumped charge and spin are simply related by: $Q/e=2S_z/\hbar$. In contrast to $h_N<h_c$ case, we found $S_z$ is quantized to be $\hbar/2$ at $\alpha=0$. 

The main physics of the above pumping process at $T=0$ can be captured by considering an effective model, where the metallic chain is coupled to a Kramers pair of MZMs, with their spin polarization direction being related to the $\bm{d}$-vector: $\bm{d}_k\propto(\cos \Theta \sin\Phi, \cos \Theta \cos \Phi, \sin \Theta)$, slowly varying with $\bm{B}_S(t)$. The coupling term is assumed to be:
\begin{equation}
    H_T=it[(c_{0\Uparrow}+c_{0\Uparrow}^{\dag})\gamma_{\Uparrow}+(c_{0\Downarrow}+c_{0\Downarrow}^{\dag})\gamma_{\Downarrow}],
\end{equation}
where $c_{0\Uparrow/\Downarrow}$ denote the electron annihilation operators at the rightmost site of the Lead $L$, $\gamma_{\Uparrow/\Downarrow}$ are the Majorana operators for the Kramers pair of MZMs. The coupling constant $t$ can be derived to be proportional to $t_{NS}\sqrt{\Delta_p\sin k_F}$\cite{Fu_2024}. Based on this we arrive at a quite meaningful conclusion that the reflection amplitudes merely depend on the $\bm{d}$-vector's orientation \cite{SM}, independent of that of $\bm{B}_S(t)$: if $h_N=0$, only Andreev reflections exist,
\begin{equation}
\begin{aligned}
   \bm{r}_{he}(E=0) &=\begin{pmatrix}
    r_{he}^{\uparrow\uparrow} & r_{he}^{\uparrow\downarrow} \\
    r_{he}^{\uparrow\downarrow} & r_{he}^{\downarrow\downarrow}
\end{pmatrix}=-\begin{pmatrix}
   \cos \Theta e^{-i\Phi} & i\sin \Theta \\
    i\sin \Theta & \cos \Theta e^{i\Phi}
\end{pmatrix} \\
&= -\sigma_y(\hat{\bm{d}}_k\cdot\boldsymbol{\sigma}),
\end{aligned}
\end{equation}
while if $h_N>h_c$, spin-up polarized incident electrons would be either totally normally reflected or totally Andreev reflected,
\begin{equation}
    \begin{aligned}
    \begin{cases}   
    r^{\uparrow\uparrow}_{he}(E=0)=-e^{-i\Phi}, &\Theta=0\\  
    r^{\uparrow\uparrow}_{ee}(E=0)=-1, &\Theta\ne0
    \end{cases}
    \end{aligned}.
\end{equation} 
Over one pumping cycle, the $\bm{d}$-vector would precess around $\bm{z}$ with nutation. While the net change of the nutation angle $\pi/2-\Theta$ is zero, the precession angle $\Phi$ advances $2\pi$, indicating only the scattered states with $\Phi$-dependent amplitudes contribute to pumping. For simplicity, to understand Fig.\ref{fig1}, consider a small $B_0$, $\theta_B$ is thus relatively fixed at $\pi/2$, which means $\Theta\approx-\alpha$. Thus according to Eq. (\ref{eq3}), if $h_N=0$, the pumped $\uparrow$-holes and $\downarrow$-holes are found to cancel each other out, resulting in no charge pumping but a spin pumping with magnitude $\hbar\cos^2\alpha$, while if $h_N>h_c$, when $\alpha=0$ the pumped $\uparrow$-hole induces the quantized pumped charge and spin: $Q/e=2S_z/\hbar=1$, and when $\alpha\ne0$ no net charge or spin is pumped due to the absence of $\Phi$-dependent reflections.

\section{Temperature and MZM-induced interference effects} 
\label{sec3}
\begin{figure}[ht]
  \begin{center}
	\includegraphics[width=8.5cm,height=10.4cm]{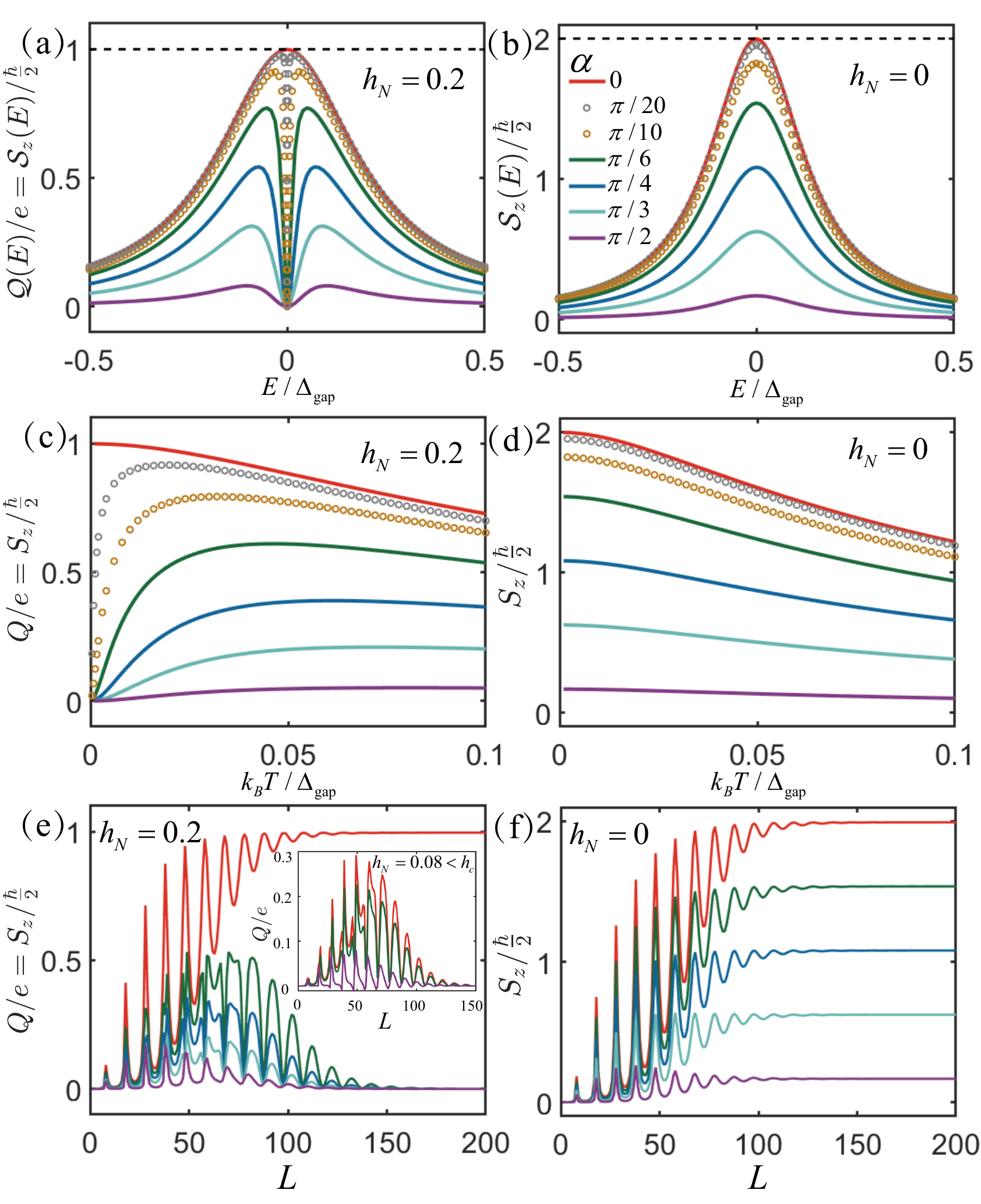}	
  \end{center}
  \vspace{-0.4cm}
	\caption{Temperature and MZM-induced interference effects of the pumping process in a $p$-wave SC, where the left (right) panels are for the cases of Lead $L$ being fully spin-polarized (unpolarized), always obeying $Q/e=2S_z/\hbar$ ($\mathcal{Q}=0$). Distribution functions (a) $\mathcal{Q}(E)$ and (b) $\mathcal{S}_z(E)$, where $\Delta_{\text{gap}}=\Delta_p\sin k_F$. $Q$ and $S_z$ as functions of $k_B T$ ((c)-(d)) or length $L$ ((e)-(f)). Parameters: $\mu=-1.9$, $L=1000$, $\Delta_p=0.02$ in (a)-(d), while $k_B T\to 0$, $\Delta_p=0.08$ in (e)-(f).
  } 
  \label{fig2}
\end{figure}

At low temperatures, only energies near the resonance $E=0$ are relevant. To demonstrate the essential physics, we only focus on the two extreme cases of $h_N>h_c$ and $h_N=0$. The numerical results of distribution functions $\mathcal{Q}(E)$ and $\mathcal{S}_z(E)$ are shown in Figs. \ref{fig2}(a)-(b). When $h_N>h_c$ and $\alpha\ne 0$, the pumped charge or spin is zero at $E=0$ due to the $\Phi$-independent total normal reflection, while near $E=0$ they behave like $aE^2$ with the coefficient $a$ predicted by the effective model to be nearly proportional to $1/\alpha^4$ for a small $\alpha$\cite{SM}, giving rise to a characteristic double-peak structure with valley width $\propto \alpha^2$ and peak value approaching $1$ as $\alpha\rightarrow0$. This is in contrast to $\mathcal{S}_z(E)$ in $h_N=0$ case, where a single peak around $E=0$ forms with its peak value approaching $2$ as $\alpha\rightarrow0$. This also leads to the fact that the pumped charge or spin as function of $k_B T$ shows a hump-like peak for $h_N>h_c$ and small $\alpha$, while the pumped spin decreases monotonically with $T$ for $h_N=0$, as shown in Figs. \ref{fig2}(c)-(d). 

The interference between MZMs at two interfaces\cite{PhysRevB.86.220506,PhysRevB.97.155425,PhysRevB.104.L020501,PhysRevB.110.115417} would become strong and may significantly influence the pumping process when $L$ is less than or of the same order of $l_M$, with $l_M\approx\hbar v_F/(\Delta_p\sin k_F)=2/\Delta_p$. As shown by open circles in Figs. \ref{fig1}(a)-(b), the numerical results of $Q$ and $S_z$ for length $L=300\approx 3l_M$ are presented, where the pumped charge becomes finite and monotonically increases with $h_N$, reaching a saturation value when $h_N>h_c$, in contrast to cases of $L \gg l_M$. In Figs. \ref{fig2}(e),(f) we further show $Q$ and $S_z$ as functions of $L$. When $L$ becomes comparable with $l_M$, the pumping shows strong oscillating behavior, with a quasi-period $\Delta L=\pi/k_F$. Furthermore, we also found that regardless of the detailed values of $h_N$, $L$ or $\alpha$, $\uparrow$-electrons and $\downarrow$-holes are always pumped in, while $\downarrow$-electrons and $\uparrow$-holes are always pumped out, indicating the pumped current is always of $100\%$ spin polarization. 

\section{Pumping in a mixed $s+p$-wave SC}
\label{sec4}
\begin{figure}[ht]
  \begin{center}
	\includegraphics[width=8.5cm,height=6.9cm]{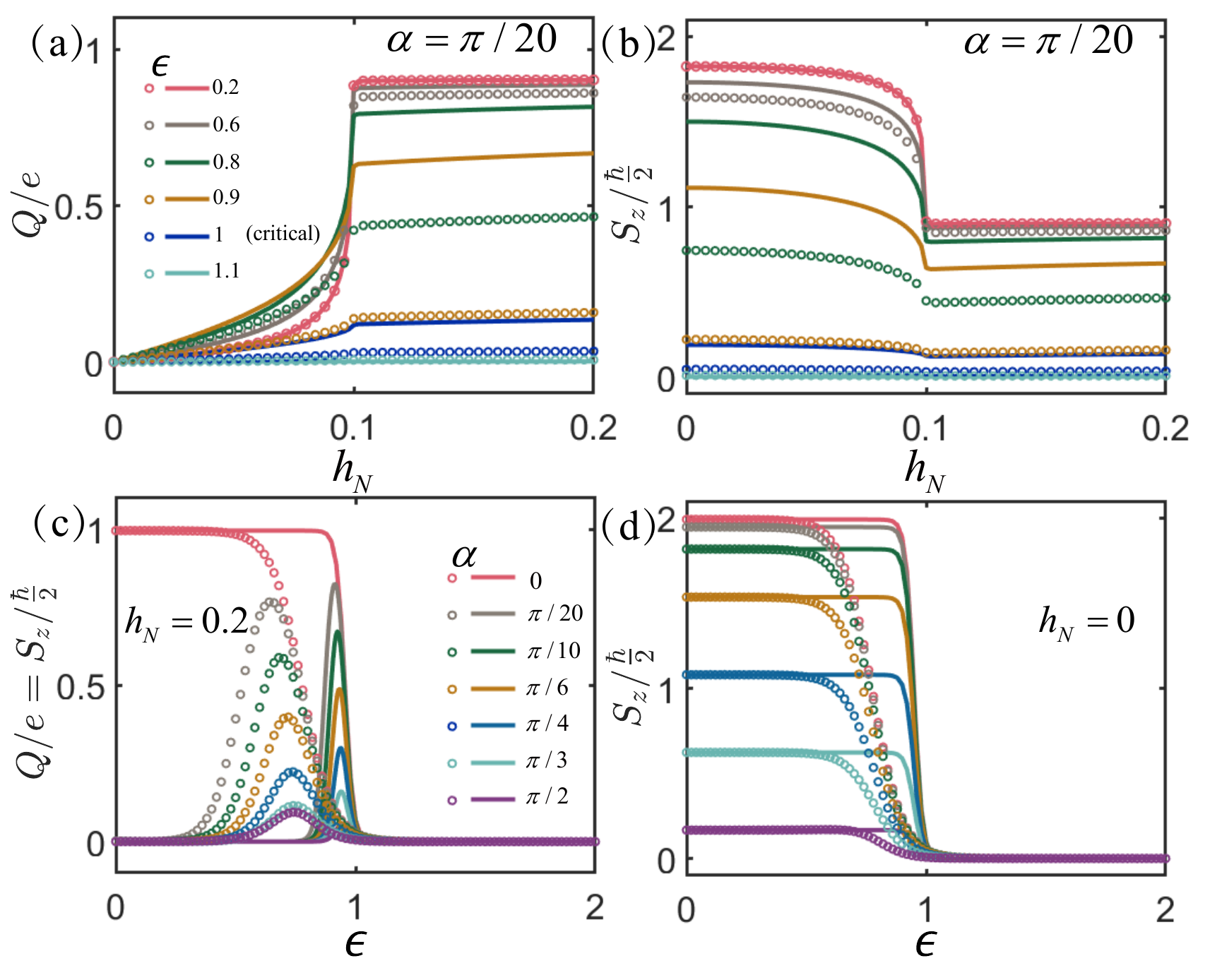}	
  \end{center}
  \vspace{-0.4cm}
	\caption{Charge and spin pumping in a mixed $s+p$-wave SC. (a) $Q$ and (b) $S_z$ as functions of $h_N$ at $k_BT/\Delta_s^c=0.02$ for different ratio $\epsilon$ of the pairing components, where $\epsilon=\Delta_s/\Delta_s^c$. (c)-(d): $Q$ and $S_z$ as functions of $\epsilon$ at $T=0$. Here $L=3000$ (800) for the solid lines (open circles), and other parameters are the same to Fig .\ref{fig1}.
  } 
  \label{fig3}
\end{figure}

In noncentrosymmetric SCs, the triplet $p$-wave pairing is typically mixed with $s$-wave pairing. This mixing effect can exhibit interesting behavior relevant to topological phase transition in a periodically pumping process and can be taken into account by simply considering an additional pairing term $\Delta_s c_{k\uparrow}^{\dag}c_{-k\downarrow}^{\dag}$ in Eq.(\ref{eq1}), with $\Delta_s$ the on-site $s$-wave pairing potential. In the weak-pairing limit ($\Delta_p, \Delta_s \ll 1$), the SC is topologically nontrivial (trivial) if the energy gap $\Delta_{\text{eff}}=\Delta_p \sin k_F-\Delta_s$ is positive (negative). As long as $p$-wave pairing is dominant ($\Delta_s<\Delta_s^c=\Delta_p \sin k_F$), both $Q$ and $S_z$ respectively exhibit monotonic increasing and decreasing behaviors as functions of $h_N$ ($h_N<h_c$), as shown in Figs. \ref{fig3}(a)-(b). When $s$-wave pairing is dominant, the pumping is quickly reduced to zero, implying that the existence of the MZMs is the key factor of a finite pumping. Because of finiteness of $L$, topological phase transition occurs relatively continuously as parameter changes. This is understood by noting the fact that $l_M\approx \hbar v_F/\left|\Delta_{\text{eff}}\right|$, so as long as variation of $\Delta_s$ makes $l_M$ comparable with $L$, interference effect would become strong enough to exhibit non-negligible size effect. This would also lead to a charge pumping resonance phenomenon at $T=0$ for $h_N>h_c$, which is demonstrated in Fig. \ref{fig3}(c), where near the critical value $\Delta_s^c$ of the phase transition, the pumped charge or spin forms sharp peak, while the pumped spin for $h_N=0$ case changes abruptly from a finite value to zero, exhibiting a step-like structure (Fig. \ref{fig3}(d)), with both the peak width and transition width being proportional to $1/L$. 

\section{Pumping in a fully spin-polarized $p$-wave SC}
\label{sec5}
\begin{figure}[ht]
  \begin{center}
	\includegraphics[width=8.5cm,height=6.8cm]{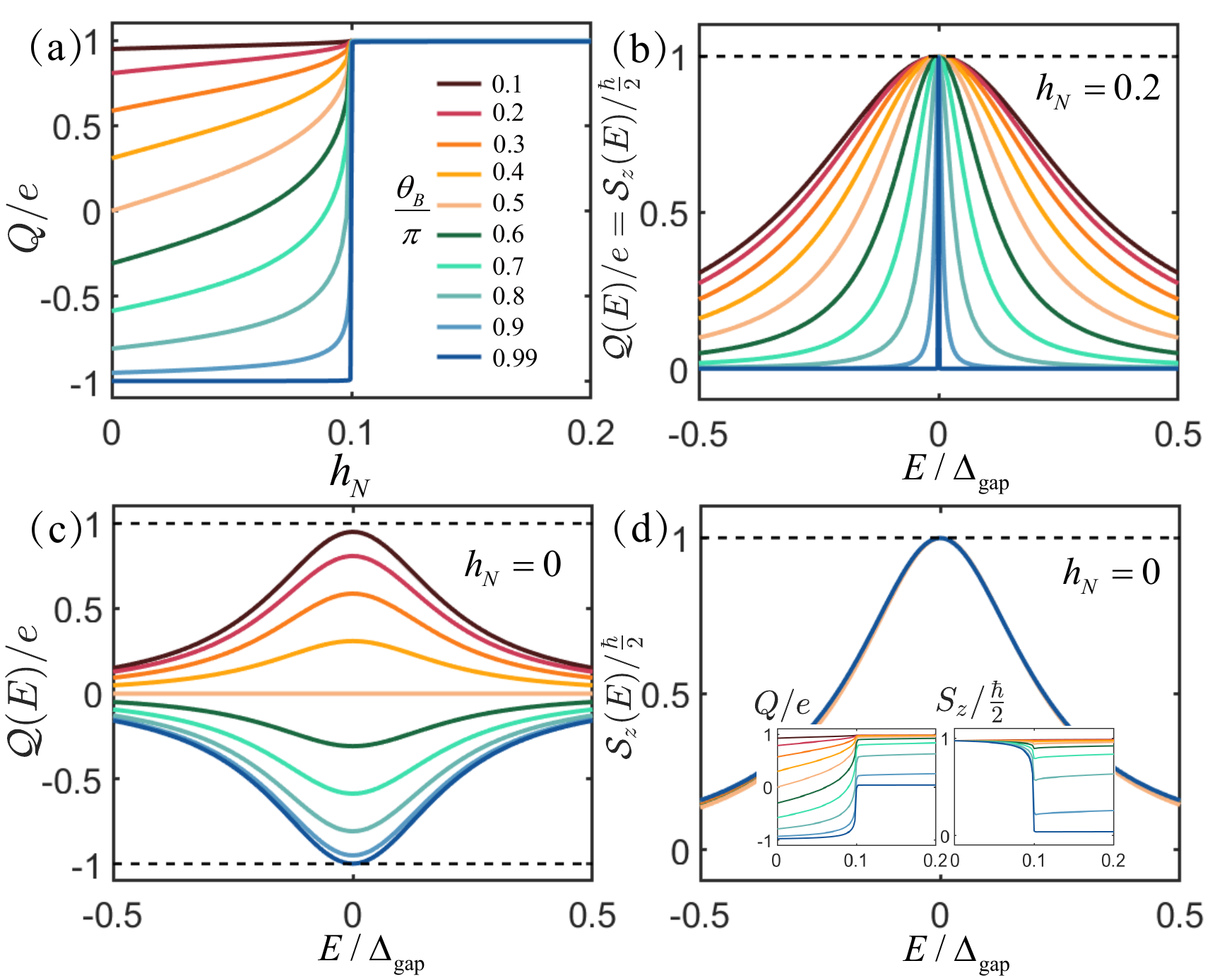}	
  \end{center}
  \vspace{-0.4cm}
	\caption{Pumping in a spin-polarized $p$-wave SC. (a) $Q$ as functions of $h_N$ for different fixed $\theta_B$ at $T=0$, while $S_z\equiv\hbar/2$. Distribution functions $\mathcal{Q}(E)$ and $\mathcal{S}_z(E)$ for (b) $h_N>h_c$ and (c)-(d) $h_N=0$. The inset in (d) gives the pumped charge and spin as functions of $h_N$ at $k_BT/(\Delta_p\sin k_F)=0.02$. Here $B_0$ varies but $B_{S}$ is fixed to be $0.2$, $L=1000$, $\mu=-1.9$, $\Delta_p=0.02$, and $t_{NS}=-0.6$.
  } \label{fig4}
\end{figure}

We now examine the pumping phenomenon for a particular $p$-wave pairing state where the magnetic field in SC is assumed to be so strong that all electrons are spin-polarized along the magnetic field and the corresponding $p$-wave pairing is described by: $i\Delta_p \sin k \ c_{k\Uparrow}^{\dag}c_{-k\Uparrow}^{\dag}$. This system is equivalent to a Kitaev chain\cite{Kitaev_2001}, hosting a single MZM at each end, whose spin polarization in the adiabatic limit is instantly orientated along the magnetic field. Here for simplicity, we assume $\theta_0=0$ and so the area swept by $\bm{B}_S$ is a circular cone, with $\theta_B=\tan^{-1}(1/\eta)$ the half apex angle. We found the pumped charge $Q$ starts from a finite value at $h_N=0$, gradually increasing up to a quantized plateau at $Q=e$ when $h_N>h_c$, as shown in Fig.\ref{fig4}(a), while the pumped spin is always quantized to be $\hbar/2$. When $h_N>h_c$, both distribution functions $\mathcal{Q}(E)$ and $\mathcal{S}_z(E)$ exhibit sharp peak as $\theta_B$ is approaching $\pi$,  as shown in Fig. \ref{fig4}(b). This is in contrast to the corresponding results for $h_N=0$ (Figs. \ref{fig4}(c)-(d)), where both $\mathcal{Q}(E)$ and $\mathcal{S}_z(E)$ still show peak structure, but while for the former, the peak value and peak width vary with $\theta_B$, the latter is nearly independent of $\theta_B$. Furthermore, $\mathcal{Q}(E)$ changes sign when $\theta_B>\pi/2$. These features give rise to the abrupt change of the pumped spin at a finite temperature from a quantized plateau at $\hbar/2$ to $0$ as $\theta_B\rightarrow\pi$, as shown in the inset of Fig. \ref{fig4}.

%***********************************
%************************************

The essential physics in this case can be understood by viewing this system as a metallic chain coupled to a single Majorana fermion, whose spin polarization is instantly along $\bm{B}_S$. Analogously, the coupling term in this case is assumed to be: $H_T=it(c_{0\Uparrow}+c_{0\Uparrow}^{\dag})\gamma_{\Uparrow}$. From this effective model, we can deduce the reflection amplitudes, which only depend on the direction of the magnetic field: $\hat{\bm{B}}_S=(\sin \theta_B\cos\phi_B,\sin\theta_B\sin\phi_B,\cos\theta_B)$, with $\theta_B$ fixed and $\phi_B=\omega t$. When $h_N=0$, we have:
\begin{equation}
    \bm{r}_{ee}(E=0)=\begin{pmatrix}
    r_{ee}^{\uparrow\uparrow} & r_{ee}^{\uparrow\downarrow} \\
    r_{ee}^{\downarrow\uparrow} & r_{ee}^{\downarrow\downarrow}
\end{pmatrix}
=\frac{e^{2ik_F}}{2}(-1+\hat{\bm{B}}_S\cdot\boldsymbol{\sigma}),
\end{equation}
and
\begin{equation}
\bm{r}_{he}(E=0)=-\frac{1}{2}\begin{pmatrix}
    (1+\cos \theta_B)e^{i\phi_B} & \sin \theta_B \\
    \sin \theta_B & (1-\cos \theta_B)e^{-i\phi_B}
\end{pmatrix}.
\end{equation}
According to Eq. (\ref{eq3}), only $\phi_B$-dependent reflection amplitudes contribute to the pumping. So the pumped charge and spin predicted by the effective model is $\mathcal{Q}(E=0)/e=\cos{\theta_B}$ and $2\mathcal{S}_z(E=0)/\hbar=1$, agreeing very well with Figs. \ref{fig4}(c)-(d). When $h_N>h_c$, only spin-up propagating modes exist, and the two nonzero reflection amplitudes are:
\begin{equation}
    \begin{aligned}
    \begin{cases}
        r^{\uparrow\uparrow}_{he}(E=0)=-e^{i\phi_B}, &\theta_B\ne\pi\\
        r^{\uparrow\uparrow}_{ee}(E=0)=-e^{2ik_F}, &\theta_B=\pi
    \end{cases}
    \end{aligned},
\end{equation}
indicating the quantized spin and charge: $\mathcal{Q}(E=0)/e=2\mathcal{S}_z(E=0)/\hbar=1$, consistent with Fig. \ref{fig4}(b). The peak width can also be predicted to be proportional to $(\theta_B-\pi)^2$\cite{SM}, as $\theta_B\rightarrow\pi$.

\section{Further discussions and conclusions}
\label{sec6}
All our discussions are based on Eq. (\ref{eq1}), which is a toy model. We now examine a more realistic effective $p$-wave model: Starting from a nanowire with Rashba spin-orbit interaction $\lambda_{\text{R}}\sin k\sigma_y$, consider the extended nearest-neighbor $s$-wave pairing $\Delta \cos k\ \tau_y\sigma_y$ induced by the proximity effect. This model was first introduced in Ref. \cite{PhysRevLett.111.056402} to achieve a time-reversal invariant topological SC in a Rashba nanowire in proximity to an $s_\pm$-wave iron-based SC. The spin-orbit interaction induces band splitting, resulting in two sets of Fermi surfaces: $\pm k_F^1$ and $\pm k_F^2$, and if the product of the effective pairings at the Fermi surfaces obeys: $\Delta_1\Delta_2<0$, the nanowire effectively becomes a topologically nontrivial SC, where $\Delta_i=\Delta \cos k_F^i$. Without loss of generality, we assume $\Delta_1$$>$$-\Delta_2$$>0$. The pairing term at Fermi surfaces becomes: $\Delta_1c_{k_F^1\Uparrow}^{\dag}c_{-k_F^1\Downarrow}^{\dag}-\Delta_2c_{k_F^2\Downarrow}^{\dag}c_{-k_F^2\Uparrow}^{\dag}$, where $\Uparrow$$(\Downarrow)$ denotes orientation along (against) $\bm{y}$. By treating the distinct Fermi surfaces as being the same at $\pm k_F$, an effective $p$-wave dominant mixed $s+p$-wave pairing is then formed and can be expressed as $\Delta_p^{\text{eff}}(c_{k\Uparrow}^{\dag}c_{-k\Downarrow}^{\dag}+c_{k\Downarrow}^{\dag}c_{-k\Uparrow}^{\dag})+\Delta_s^{\text{eff}}(c_{k\Uparrow}^{\dag}c_{-k\Downarrow}^{\dag}-c_{k\Downarrow}^{\dag}c_{-k\Uparrow}^{\dag})$, where $\Delta_{s/p}^{\text{eff}}=\ (\Delta_1\pm\Delta_2)/2$. Due to time-reversal symmetry, this chain also hosts a Karamers pair of MZMs at each end or interface. Thus an effective $p$-wave $\bm{d}$-vector can be well defined, whose direction is same to the that of the Rashba interaction and so is always perpendicular to the nanowire axis. When adiabatically rotating gate voltage above the chain, the direction of Rashba interaction and then the effective $p$-wave $\bm{d}$-vector would rotate periodically around the chain \cite{PhysRevLett.78.1335,PhysRevB.94.035444}. The pumping results over one cycle have been confirmed to be similar to those for $\alpha=0$ in Fig. \ref{fig1}. Therefore, our results of the charge and spin pumping based on $p$-wave superconductors can be extended to general time-reversal invariant topological SCs and these phenomena could serve as transport signatures of the MZMs.
\section{Acknowledgment}
J.J.F. thanks Shu-tong Guan and Dao-he Ma for helpful discussions. This work was supported by NSFC under Grant No.~11874202, and Innovation Program for Quantum Science and Technology under Grant No.2024ZD0300101.

\bibliographystyle{unsrt}
\bibliography{main}

\clearpage
\renewcommand{\thefigure}{S\arabic{figure}}
\renewcommand{\theequation}{S\arabic{equation}}
\setcounter{figure}{0}
\setcounter{equation}{0}

\widetext
\begin{center}
\textbf{\large Supplemental Material for \\$\pmb{d}$-vector precession induced pumping in topological $p$-wave superconductors} 

\bigskip

Jun-Jie Fu$^1$ and Jin An$^{1,2,*}$

$^{\it{1}}$\textit{National Laboratory of Solid State Microstructures, Department of Physics, Nanjing University, Nanjing 210093, China}\\
$^{\it{2}}$\textit{Collaborative Innovation Center of Advanced Microstructures, Nanjing University, Nanjing 210093, China}
\end{center}

\section*{I. Coupled Majorana Fermion in the quantum transport of a metallic chain: spinless case}
\begin{figure}[ht]
  \begin{center}
	\includegraphics[width=10cm,height=3.33cm]{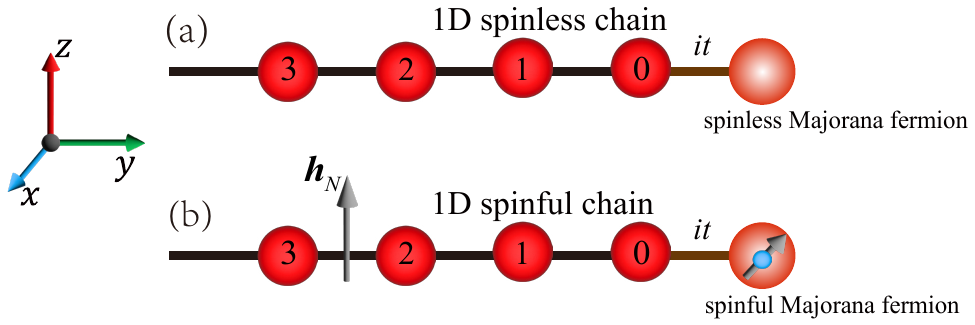}	
  \end{center}
  \vspace{-0.4cm}
	\caption{Effective model of the metallic chain coupled with a topological superconductor, where the normal metallic lead is (a) spinless, or (b) spinful. Here $it$ is the coupling constant between the lead and the Majorana fermion, and $\bm{h}_N$ is the Zeeman field applied in the normal lead.
  } \label{figa1}
\end{figure}

Here in this section, we give an analytical derivation of the quantum transport in the metallic spinless chain coupled to a Kitaev superconductor using the effective model shown in Fig. \ref{figa1}, where the Kitaev superconductor is treated effectively as an isolated Majorana fermion. From this effective model, we shall derive the essential result that an incident electron would be totally reflected as a hole at resonance energy $E=0$.     

The lead, as a normal metallic chain, can be described as:
\begin{equation}
   \mathcal{H}_N(k)=(-2\cos k-\mu)\tau_z,
\end{equation}
where $\bm{\tau}=(\tau_x,\tau_y,\tau_z)$ are Pauli matrices acting in Nambu space with the basis $(c_{k},c^{\dag}_{-k})^T$. It is assumed that the Majorana fermion is coupled to the lead only via the end site of the latter. The coupling term can be given by:
\begin{equation}
    H_T=it(c_0+c_0^{\dag})\gamma,
    \label{eq2}
\end{equation}
where $it$ is the coupling constant. Here $c_0$ denotes the electron annihilation operator at the rightmost site of the lead and $\gamma$ is the Majorana fermion operator, which in the fermion representation can be expressed as $\gamma=f+f^{\dag}$. This coupling term can be further written as:
\begin{equation}
    H_T = it(c_0^{\dag},c_0)
    \begin{pmatrix}
    1 & 1 \\
    1 & 1
   \end{pmatrix}
   \begin{pmatrix}
    f\\
    f^{\dag}
   \end{pmatrix}.
\end{equation}
Thus in the particle-hole Nambu representation this effective model is converted to be one that the end site of the lead is coupled to a new lattice site via the following effective hopping matrix:
\begin{equation}
    T_{\text{eff}}=it(1+\tau_x).
\end{equation}

Now treat the end lattice site of the lead as the scattering region and suppose that an electron with energy $E$ is incident from the other end of the lead. The self-energy contributed from the new extra lattice site is:
\begin{equation}
    \Sigma_R^r=T_{\text{eff}}\frac{1}{E^+}T^{\dag}_{\text{eff}}=\frac{2t^2}{E^+}(1+\tau_x),
\end{equation}
where $E^+=E+i0^+$. By comparison with the self-energy of the original lattice model~\cite{Fu_2024}, one can derive that $t\propto t_{NS}\sqrt{\Delta/E_F}$, where $t_{NS}$ is the interface hopping integral, $\Delta$ is the pairing order parameter and $E_F$ is the Fermi energy. In addition to this, the lead itself would give a standard contribution to the self-energy by the surface Green function:
\begin{equation}
    \Sigma_L^r=g_L^r=\begin{pmatrix}
    -e^{ik_e} & 0 \\
    0 & e^{-ik_h}
\end{pmatrix},
\end{equation}
where $g_L^r$ is the lead's surface Green function, obeying $E+\mu \tau_z-g_L^r=(g_L^r)^{-1}$. $k_e$ ($k_h$) is the wave vector of the incident electron (reflected hole) in the lead, satisfying $E = -2 \cos k_e-\mu=2\cos k_h+\mu$. The retarded Green function of the scattering region can then be given by:
\begin{equation}
G^r(E) =(E+\mu \tau_z-\Sigma_L^r-\Sigma_R^r)^{-1}= [\frac{1}{2}(e^{ik_h}-e^{-ik_e})-\frac{1}{2}(e^{ik_h}+e^{-ik_e})\tau_z-\frac{2t^2}{E^+}(1+\tau_x)]^{-1}.
\label{eq7}
\end{equation}
At $E = 0$, the wave vectors satisfy $k_e=k_h = k_F$, with $k_F$ the Fermi wave vector. So the retarded Green function can be simplified as:
\begin{equation}
    G^r(E=0)=\frac{1}{2i\sin k_F}(1-\tau_x).
    \label{eq8}
\end{equation}
Using the Fisher-Lee relation~\cite{PhysRevB.23.6851}, the reflection entries of the spinless scattering matrix can be obtained by:
\begin{equation}
   r_{\alpha\beta}(E)=-\delta_{\alpha\beta}+i[\Gamma^L_{\alpha\alpha}(E)]^{1/2}G_{\alpha\beta}^r(E)[\Gamma^L_{\beta\beta}(E)]^{1/2}.
    \label{eq9}
\end{equation}
Here $r_{\alpha\beta}$ represents the reflection amplitude where the incident particle with state $\beta$ is reflected as $\alpha$ with $\alpha,\beta\in \{e,h\}$. $\Gamma^L_{ee}$ and $\Gamma^L_{hh}$ are the line-width functions of the lead, which are proportional to the group velocities of the propagating electron and hole respectively, sharing the same value at $E=0$: $\Gamma^L_{ee}(E=0)=\Gamma^L_{hh}(E=0)=v_F=2\sin k_F$. The local Andreev reflection amplitude $r_{he}$, where the incident electron is reflected as a hole, and the normal reflection amplitude $r_{ee}$, where the incident electron is still reflected as an electron, can be deduced as:
\begin{equation}  
        r_{he}(E=0) = -1,\ 
        r_{ee}(E=0)=0.
\end{equation}
The Andreev and normal reflection coefficients $R_A$ and $R_N$ can then be obtained:
\begin{equation}
        R_A(E=0) = |r_{he}|^2=1,\ 
    R_N(E=0)=|r_{ee}|^2=0.
\end{equation}
This indicates the resonant complete Andreev reflection. Namely, at $E=0$ , an incident electron would be completely reflected as a hole. Obviously this phenomenon is induced by the existence of the coupled Majorana fermion and this is also in accordance with the previous results in Ref. ~\cite{PhysRevLett.103.237001}.

The transport properties near the resonance can also be obtained. According to Eq. (\ref{eq7}), we generally have: 
\begin{equation} 
\begin{aligned}
    &G^r_{he}(E)=\frac{2t^2/E}{-e^{i(k_h-k_e)}-2(e^{ik_h}-e^{-ik_e})t^2/E},\\
    &G^r_{ee}(E)=\frac{e^{ik_h}-2t^2/E}{-e^{i(k_h-k_e)}-2(e^{ik_h}-e^{-ik_e})t^2/E}.
\end{aligned}
\label{eq12}
\end{equation}
The line-width functions take the values $\Gamma_{ee}^L = 2 \sin k_e$, $\Gamma_{hh}^L = 2 \sin k_h$. As energy $E$ of the incident electron approaches $0$, namely, $E\rightarrow0$, by making use of the approximation $k_e-k_F=k_F-k_h\approx E/v_F=E/(2\sin k_F)$, the Andreev and normal reflection amplitudes as well as reflection coefficients can be deduced as: 
\begin{equation}
   \begin{aligned}
    &r_{he}(E)=i[\Gamma^L_{hh}]^{1/2}G^r_{he}[\Gamma^L_{ee}]^{1/2}\approx-1-\frac{i(2t^2+1)}{2v_Ft^2}E+\frac{v_F^2(1+2t^4)+8t^4}{4v_F^4t^4}E^2,\\
    &r_{ee}(E)=-1+i[\Gamma^L_{ee}]^{1/2}G^r_{ee}[\Gamma^L_{ee}]^{1/2}\approx \frac{4t^2e^{ik_F}+iv_Fe^{2ik_F}}{2v_F^2t^2}E,\\
    &R_A(E) = |r_{he}(E)|^2\approx1-\frac{1}{v_F^2}(\frac{4}{v_F^2}+\frac{1-4t^2}{4t^4})E^2,\\
    &R_N(E) = 1-R_A(E).
   \end{aligned}
   \label{eq13}
\end{equation}
Since $it$ is an effective coupling constant, it varies as system parameters change. If $t$ becomes sufficiently small, i.e., $t\ll1$(in the weak-pairing limit of $\Delta/E_F\ll1$ as an example), a small deviation from $E=0$ is expected to cause a finite Andreev reflection:
\begin{equation}
    R_A(E) \approx 1-\frac{E^2}{4v_F^2t^4}.  
\end{equation}
Accordingly, in this case the reflection amplitudes take the following forms:
\begin{equation}
\begin{aligned}
    &r_{he}(E) \approx -1-i\frac{E}{2v_Ft^2}+\frac{E^2}{4v_F^2t^4},\\
    &r_{ee}(E) \approx i\frac{e^{2ik_F}E}{2v_Ft^2}.
\end{aligned}   
\end{equation}

\section*{II.Coupled Majorana Fermion in the quantum transport of a metallic chain: spinful case}

We continue in this section to discuss the quantum transport in a spinful metallic chain, which is coupled to a fully spin-polarized $p$-wave superconductor. The coupling to the superconductor is described here as that to a spin-definite Majorana fermion described as $\gamma_{\Uparrow}=f_{\Uparrow}+f_{\Uparrow}^{\dag}$. $f_{\Uparrow}$ is a fermion operator which is spin directed along the spin-polarization direction of the superconductor. This is because in the pumping process, the spin polarization is assumed to vary adiabatically, the spin direction of $f_{\Uparrow}$ is expected to instantly follow it. For the normal metallic chain, a Zeeman field $\bm{h}_N$ are also taken into account, the direction of which is assumed to be always along $\boldsymbol{z}$. This situation is shown schematically in Fig. \ref{figa1}(b). As before, the coupling between the lead and the Majorana fermion is given by:
\begin{equation}
 H_T=it(c_{0\Uparrow}+c_{0\Uparrow}^{\dag})\gamma_{\Uparrow},
    \label{eq16}
\end{equation}
where only the spin-$\Uparrow$ electron at the end site of the lead is assumed to couple the Majorana fermion. 

In the absence of the Zeeman field $\bm{h}_N$, the situation is rather simple. There are two independent propagation modes in the lead, which can be chosen to be $\Uparrow$ and $\Downarrow$, respectively. Eq. (\ref{eq16}) means only the spin-$\Uparrow$ mode is coupled to the Majorana fermion, and its coupling form is similar to Eq. (\ref{eq2}). Hence the incident spin-$\Uparrow$ electrons will be totally reflected as a spin-$\Uparrow$ hole with the amplitude $r_{he}^{\Uparrow\Uparrow}(E=0)=-1$. For the incident spin-$\Downarrow$ mode, which has no coupling to the Majorana fermion, its retarded Green function for the scattering region(the end site of the lead) can be written as:
\begin{equation}
    G_{\Downarrow\Downarrow}^r(E) =(E+\mu \tau_z-\Sigma_L^r)^{-1}=\begin{pmatrix}
    -e^{ik_e} & 0 \\
    0 & e^{-ik_h}
\end{pmatrix}.
\end{equation}
Then according to Eq. (\ref{eq9}), one acquires $r_{ee}^{\Downarrow\Downarrow}(E)=-e^{2ik_e}$,  and further $r_{ee}^{\Downarrow\Downarrow}(E=0)=-e^{2ik_F}$. 

If the two independent propagation modes in the lead are chosen to be the standard $\uparrow$ and $\downarrow$, which is parallel or opposite to the fixed $\boldsymbol{z}$ direction, one can obtain the reflection amplitudes for both modes by a spin rotation transformation. Namely, one can introduce $(c_{\Uparrow}^{\dag},c_{\Downarrow}^{\dag})=(c_{\uparrow}^{\dag},c_{\downarrow}^{\dag})U$, where $U=U_z(\phi_B)U_y(\theta_B)$ with $U_{\bm{n}}(\theta)=\exp (-i\frac{\theta}{2}\bm{\sigma}\cdot \bm{n})$ representing a spin rotation by $\theta$ around $\bm{n}$. Here $\theta_B$ and $\phi_B$ are the spherical Euler angles of the spin-polarization direction in the superconductor, which is : $(\sin \theta_B \cos \phi_B,\sin \theta_B \sin \phi_B,\cos \theta_B)$. The particle$\rightarrow$particle and particle$\rightarrow$hole reflection matrices in this representation are then given by:
\begin{equation}
    \bm{r}_{ee}(E=0)=\begin{pmatrix}
    r_{ee}^{\uparrow\uparrow} & r_{ee}^{\uparrow\downarrow} \\
    r_{ee}^{\downarrow\uparrow} & r_{ee}^{\downarrow\downarrow}
\end{pmatrix}=U\begin{pmatrix}
    0 & 0 \\
    0 & -e^{2ik_F}
\end{pmatrix}U^{\dag}\\
=\frac{-e^{2ik_F}}{2}\begin{pmatrix}
    1-\cos \theta_B & -\sin \theta_B e^{-i\phi_B} \\
    -\sin \theta_B e^{i\phi_B} & 1+\cos \theta_B
\end{pmatrix},
\end{equation}
and
\begin{equation}
\bm{r}_{he}(E=0)=\begin{pmatrix}
    r_{he}^{\uparrow\uparrow} & r_{he}^{\uparrow\downarrow} \\
    r_{he}^{\downarrow\uparrow} & r_{he}^{\downarrow\downarrow}
\end{pmatrix}=U^{*}\begin{pmatrix}
    -1 & 0 \\
    0 & 0
\end{pmatrix}U^{\dag}\\
=-\frac{1}{2}\begin{pmatrix}
    (1+\cos \theta_B)e^{i\phi_B} & \sin \theta_B \\
    \sin \theta_B & (1-\cos \theta_B)e^{-i\phi_B}
\end{pmatrix}.
\end{equation}

Near the resonance, the Andreev 
 and normal reflection amplitudes for the incident spin-$\Uparrow$ electron can be given from Eq. (\ref{eq13}):
\begin{equation}
   \begin{aligned}
    &r_{he}^{\Uparrow\Uparrow}(E)\approx-1-\frac{i(2t^2+1)}{2v_Ft^2}E+\frac{v_F^2(1+2t^4)+8t^4}{4v_F^4t^4}E^2,\\
    &r_{ee}^{\Uparrow\Uparrow}(E)\approx \frac{4t^2e^{ik_F}+iv_Fe^{2ik_F}}{2v_F^2t^2}E+\frac{(1+2t^2)(e^{2ik_F}-1)(e^{2ik_F}+4t^2-1)}{4v_F^4t^4}E^2. 
   \end{aligned}
\end{equation}
For the incident spin-$\Downarrow$ electron, there is only normal reflection with amplitude: $r_{ee}^{\Downarrow\Downarrow}(E)=-e^{2ik_e}\approx -e^{2ik_F}(1+2iE/v_F-2E^2/v_F^2)$. The reflection matrices in the standard representation are given by:
\begin{equation}
\begin{aligned}
    \bm{r}_{ee}(E)
&\approx\frac{-e^{2ik_F}}{2}(1+2i\frac{E}{v_F}-\frac{2E^2}{v_F^2})\begin{pmatrix}
    1-\cos \theta_B & -\sin \theta_B e^{-i\phi_B} \\
    -\sin \theta_B e^{i\phi_B} & 1+\cos \theta_B
\end{pmatrix}\\
&+(\frac{(4t^2e^{ik_F}+iv_Fe^{2ik_F})}{4v_F^2t^2}E+\frac{(1+2t^2)(e^{2ik_F}-1)(e^{2ik_F}+4t^2-1)}{8v_F^4t^4}E^2)\begin{pmatrix}
    1+\cos \theta_B & \sin \theta_B e^{-i\phi_B} \\
    \sin \theta_B e^{i\phi_B} & 1-\cos \theta_B
\end{pmatrix},
\end{aligned} 
\end{equation}
and
\begin{equation}
\bm{r}_{he}(E)
\approx\frac{1}{2}(-1-\frac{i(2t^2+1)}{2v_Ft^2}E+\frac{v_F^2(1+2t^4)+8t^4}{4v_F^4t^4}E^2)\begin{pmatrix}
    (1+\cos \theta_B)e^{i\phi_B} & \sin \theta_B \\
    \sin \theta_B & (1-\cos \theta_B)e^{-i\phi_B}
\end{pmatrix}.
\end{equation}

With the adiabatic variation of the magnetic field described in the main text, the spin-polarization direction of the superconductor is precessing around $\bm{z}$ in a circular cone with half apex angle $\theta_B$. The $E$-dependent charge and spin pumping with small $t$ can be obtained as:
\begin{equation}
    \begin{aligned}
        &\mathcal{Q}(E)/e=|r_{ee}^{\uparrow\downarrow}|^2-|r_{ee}^{\downarrow\uparrow}|^2+|r_{he}^{\uparrow\uparrow}|^2-|r_{he}^{\downarrow\downarrow}|^2\approx\cos \theta_B(1-\frac{1}{4v_F^2t^4}E^2),\\ 
        &\mathcal{S}_z(E)/\frac{\hbar}{2}=|r_{ee}^{\uparrow\downarrow}|^2+|r_{ee}^{\downarrow\uparrow}|^2+|r_{he}^{\uparrow\uparrow}|^2+|r_{he}^{\downarrow\downarrow}|^2\approx 1+\frac{\sin^2\theta_B\cos k_F}{2v_F^2}E-\frac{1}{4v_F^2t^4}E^2.
    \end{aligned}
\end{equation}
 
In the presence of $\bm{h}_N$, the normal lead will be spin-polarized. We consider the particular case where the lead is fully spin-polarized when $\bm{h}_N$ is sufficiently large. In this case there exists only spin-$\uparrow$ propagation mode, the contribution from spin-$\downarrow$ mode in the coupling term can be removed. So the coupling term becomes:
\begin{equation}
    H_T\longrightarrow (c_{0 \uparrow}^{\dag},c_{0\uparrow})T_{\text{eff}}\begin{pmatrix}
    f_{\uparrow} \\
    f_{\downarrow} \\
    f_{\uparrow}^{\dag}\\
    f_{\downarrow}^{\dag}
\end{pmatrix},
\end{equation}
where
\begin{equation}
    T_{\text{eff}}=it\cos \frac{\theta_B}{2}
    \begin{pmatrix}
    e^{-i\frac{\phi_B}{2}}\\
    e^{i\frac{\phi_B}{2}}
    \end{pmatrix}
    \begin{pmatrix}
    \cos \frac{\theta_B}{2}e^{i\frac{\phi_B}{2}}, & \sin \frac{\theta_B}{2}e^{-i\frac{\phi_B}{2}}, & \cos \frac{\theta_B}{2}e^{-i\frac{\phi_B}{2}}, & \sin \frac{\theta_B}{2}e^{i\frac{\phi_B}{2}}
\end{pmatrix}.
\end{equation}
When $\theta_B=\pi$, this coupling is zero and so a spin-$\uparrow$ incident electron would be totally normally reflected with the reflection amplitude $r_{ee}^{\uparrow\uparrow}=-e^{2ik_F}$. Otherwise, when $\theta_B\neq\pi$, this coupling term would induce the following self-energy:
\begin{equation}
    \Sigma^r_R=\frac{T_{\text{eff}}T_{\text{eff}}^{\dag}}{E^+}=\frac{2t^2\cos^2 \frac{\theta_B}{2}}{E^+}(1+\cos \phi_B \tau_x+\sin \phi_B \tau_y).
\end{equation}
When $E= 0$, the retarded Green function of the scattering region can be deduced similarly to Eq. (\ref{eq8}) as:
\begin{equation}
        G^r(E= 0)=\frac{1}{2i\sin k_F}(1-\cos \phi_B\tau_x-\sin \phi_B \tau_y).
\end{equation}
Then according to Eq. (\ref{eq9}), we have:
\begin{equation}
    \begin{aligned}
    \begin{cases}
         r^{\uparrow\uparrow}_{ee}(E=0)=0,
         \ \ \ \ \ \ \ \ \ \ \ \ r^{\uparrow\uparrow}_{he}(E=0)=-e^{i\phi_B},
         &\theta_B\ne\pi;\\
        r^{\uparrow\uparrow}_{ee}(E=0)=-e^{2ik_F},
         \ \ r^{\uparrow\uparrow}_{he}(E=0)=0,&\theta_B=\pi.
    \end{cases}
    \end{aligned}
\end{equation}

To obtain the reflection properties near the resonance, we start from the general forms of the retarded Green functions:
\begin{equation}
    \begin{aligned}
        &G_{he}^r(E)=\frac{2t'^2/E }{-e^{i(k_h-k_e)}-2(e^{ik_h}-e^{-ik_e})t'^2/E}e^{i\phi_B},\\
        &G_{ee}^r(E)=\frac{e^{ik_h}-2t'^2/E }{-e^{i(k_h-k_e)}-2(e^{ik_h}-e^{-ik_e})t'^2/E},
    \end{aligned}
    \label{eq29}
\end{equation}
where $t'=t\cos \frac{\theta_B}{2}$. They share the similar forms to Eq. (\ref{eq12}). So as $E\rightarrow0$, by a analogous derivation, we have the following Andreev and normal reflection amplitudes, as well as the reflection coefficients:
\begin{equation}
    \begin{aligned}
        &r_{he}(E)\approx(-1-\frac{i(2t'^2+1)}{2v_Ft'^2}E+\frac{v_F^2(1+2t'^4)+8t'^4}{4v_F^4t'^4}E^2)e^{i\phi_B},\\
        &r_{ee}(E)\approx \frac{4t'^2e^{ik_F}+iv_Fe^{2ik_F}}{2v_F^2t'^2}E,\\
        &R_A(E) = |r_{he}(E)|^2\approx1-\frac{1}{v_F^2}(\frac{4}{v_F^2}+\frac{1-4t'^2}{4t'^4})E^2,\\
    &R_N(E) = 1-R_A(E).
    \end{aligned}
\end{equation}
If $\theta_B\to \pi$, which means the spin-polarization direction of the superconductor is nearly opposite to $\bm{h}_N$, the effective coupling constant $t'=t\cos \frac{\theta_B}{2}$ is approaching $0$. The reflection amplitudes and coefficients can be further simplified as:
\begin{equation}
    \begin{aligned}
        &r_{he}(E)\approx (-1-i\frac{E}{2v_Ft'^2}+\frac{E^2}{4v_F^2t'^4})e^{i\phi_B} \approx(-1-i\frac{2E}{v_Ft^2(\delta\theta_B)^2}+\frac{4E^2}{v_F^2t^4(\delta\theta_B)^4})e^{i\phi_B},\\
        &r_{ee}(E)\approx i\frac{e^{2ik_F}E}{2v_Ft'^2}\approx i\frac{2e^{2ik_F}E}{v_Ft^2(\delta\theta_B)^2},\\
        &R_A(E)\approx1-(\frac{2E}{v_Ft^2(\delta\theta_B)^2})^2,
    \end{aligned}
\end{equation}
where $\delta\theta_B=\pi-\theta_B$.

In this case, the $E$-dependent charge and spin pumping can be easily obtained as:
\begin{equation}
    \mathcal{Q}(E)/e=\mathcal{S}_z(E)/\frac{\hbar}{2}=R_A\xrightarrow{\theta_B\to\pi}1-(\frac{2E}{v_Ft^2(\delta\theta_B)^2})^2.
\end{equation}

\section*{III. Coupled Kramers pair of Majorana fermions in the quantum transport of a metallic chain}
\begin{figure}[ht]
  \begin{center}
	\includegraphics[width=10cm,height=2.5cm]{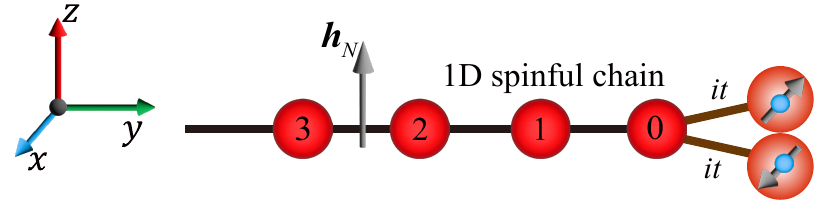}	
  \end{center}
  \vspace{-0.4cm}
	\caption{Effective model of a spinful metallic chain coupled with a time-reversal symmetric $p$-wave superconductor, where the normal lead is coupled to two Kramers degenerate Majorana fermions. Here $it$ is the coupling constant between the lead and the Majorana fermions, and $\bm{h}_N$ is the Zeeman field applied in the normal lead.
  } \label{figa2}
\end{figure}
The analysis in the above sections can now be extended to the situation where the metallic lead is coupled with a standard time-reversal symmetric $p$-wave superconductor, where both spin-parallel pairing between spin-up electrons and that between spin-down electrons equivalently coexist. Suppose that the $\bm{d}$-vector of the $p$-wave pairing is initially along $\bm{y}$, indicating that the two pairing channels share the same pairing phase. This means that at the interface between the lead and superconductor the two time-reversal related Majorana fermions can be described as $\gamma_{\uparrow}=f_{\uparrow}+f_{\uparrow}^{\dag} $ and $\gamma_{\downarrow}=f_{\downarrow}+f_{\downarrow}^{\dag} $. Thus analogously this hybrid system can be viewed as an effective model shown schematically in Fig. \ref{figa2}, where the coupling term is assumed to be:
\begin{equation}
    H_T=it[(c_{0\uparrow}+c_{0\uparrow}^{\dag})\gamma_{\uparrow}+(c_{0\downarrow}+c_{0\downarrow}^{\dag})\gamma_{\downarrow}
    ].
\end{equation}
Under an adiabatic variation of the $\bm{d}$-vector in the pumping process, the coupling term becomes:
\begin{equation}
    H_T=it[(c_{0\Uparrow}+c_{0\Uparrow}^{\dag})\gamma_{\Uparrow}+(c_{0\Downarrow}+c_{0\Downarrow}^{\dag})\gamma_{\Downarrow}
    ],
    \label{eq34}
\end{equation}
where $\Uparrow$ is along the spin-polarization direction $(-\sin \Theta \sin \Phi,-\sin \Theta \cos \Phi,\cos \Theta)$ in the $p$-wave superconductor, which is modulated adiabatically by a varying tiny magnetic field $\bm{B}_S$ in the superconductor.    
In the absence of the Zeeman field $\bm{h}_N$ in the normal lead, the two independent propagation modes which can still be chosen to be spin-$\Uparrow$ and spin-$\Downarrow$, are decoupled with each other. Thus, at the resonance energy $E=0$ an incident spin-$\Uparrow$(-$\Downarrow$) electron would be totally converted to spin-$\Uparrow$(-$\Downarrow$) hole without normal reflection. So the particle$\rightarrow$hole reflection matrix is a negative identity matrix. By introducing a spin-rotation transformation: $(c_{\Uparrow}^{\dag},c_{\Downarrow}^{\dag})=(c_{\uparrow}^{\dag},c_{\downarrow}^{\dag})U$, where $U=U_z(-\Phi)U_x(\Theta)$, at each instant, the $\bm{d}$-vector is perpendicular to the spin-polarization direction of the superconductor, and can be derived to be $\hat{\bm{d}}=(\cos \Theta \sin \Phi,\cos \Theta \cos \Phi,\sin \Theta)$. Therefore, the particle$\rightarrow$hole reflection matrix in the standard spin-$\uparrow\downarrow$ representation becomes:
\begin{equation}
       \bm{r}_{he}(E=0)=\begin{pmatrix}
    r_{he}^{\uparrow\uparrow} & r_{he}^{\uparrow\downarrow} \\
    r_{he}^{\downarrow\uparrow} & r_{he}^{\downarrow\downarrow}
\end{pmatrix}=U^{*}\begin{pmatrix}
    -1 & 0 \\
    0 & -1
\end{pmatrix}U^{\dag}\\
=-\begin{pmatrix}
    \cos \Theta\ e^{-i\Phi} &  i\sin \Theta \\
     i\sin \Theta &  \cos \Theta\ e^{i\Phi}
\end{pmatrix}.
\end{equation}
This result is independent of the final orientation of $\bm{B}_S$ but only depends on that of $\bm{d}$-vector. Actually if we further perform a spin rotation around $\bm{d}$-vector by an arbitrary angle $\beta$, $\bm{d}$-vector is left unchanged but $\bm{B}_S$ varies, and the reflection matrix $\bm{r}_{he}$ would become $U_{\bm{d}}^*(\beta)\bm{r}_{he}U_{\bm{d}}^{\dag}(\beta)$ which can be verified to be identical to $\bm{r}_{he}$.

Near the resonance, the reflection amplitudes for spin-$\Uparrow$ and spin-$\Downarrow$ propagation modes can be given from Eq. (\ref{eq13}) as ($t\ll1$):
\begin{equation}  
\begin{aligned}
    &r_{he}^{\Uparrow\Uparrow}(E)= r_{he}^{\Downarrow\Downarrow}(E)\approx-1-i\frac{E}{2v_Ft^2}+\frac{E^2}{4v_F^2t^4}, \\
    &r_{ee}^{\Uparrow\Uparrow}(E)= r_{ee}^{\Downarrow\Downarrow}(E)\approx i\frac{e^{2ik_F}}{2v_Ft^2}E. 
\end{aligned}       
\end{equation}
$\bm{r}_{he}(E)$ and $\bm{r}_{ee}(E)$ in the standard spin-$\uparrow\downarrow$ representation become:
\begin{equation}
\begin{aligned}
       &\bm{r}_{he}(E)=\begin{pmatrix}
    r_{he}^{\uparrow\uparrow} & r_{he}^{\uparrow\downarrow} \\
    r_{he}^{\downarrow\uparrow} & r_{he}^{\downarrow\downarrow}
\end{pmatrix}\approx(-1-i\frac{E}{2v_Ft^2}+\frac{E^2}{4v_F^2t^4})\begin{pmatrix}
    \cos \Theta\ e^{-i\Phi} &  i\sin \Theta \\
     i\sin \Theta &  \cos \Theta\ e^{i\Phi}
\end{pmatrix},\\
&\bm{r}_{ee}(E)=\begin{pmatrix}
    r_{ee}^{\uparrow\uparrow} & r_{ee}^{\uparrow\downarrow} \\
    r_{ee}^{\downarrow\uparrow} & r_{ee}^{\downarrow\downarrow}
\end{pmatrix}\approx i\frac{e^{2ik_F}}{2v_Ft^2}E\begin{pmatrix}
    1 &  0 \\
     0 &  1
\end{pmatrix}.
\end{aligned}
\end{equation}

When $\bm{d}$-vector varies adiabatically, following $\bm{B}_s(t)$ in the main text, the direction of $\bm{d}$-vector satisfies: $\Theta\approx -\alpha, \Phi\approx\omega t$. The $E$-dependent charge and spin pumping can be obtained as ($t\ll1$):
\begin{equation}
    \mathcal{Q}(E)/e\approx|r_{he}^{\uparrow\uparrow}|^2-|r_{he}^{\downarrow\downarrow}|^2=0,\ \mathcal{S}_z(E)/\frac{\hbar}{2}\approx|r_{he}^{\uparrow\uparrow}|^2+|r_{he}^{\downarrow\downarrow}|^2\approx 2\cos^2 \alpha(1-(\frac{E}{2v_Ft^2})^2).
\end{equation}

For the case of the fully spin-polarized metallic chain in the presence of a sufficiently large Zeeman field $\bm{h}_N$,  the spin-$\uparrow$ mode is the only propagation mode. By removing the contribution from the spin-$\downarrow$ mode, the coupling term in Eq. (\ref{eq34}) becomes: 
\begin{equation}
   H_T\longrightarrow  (c_{0\uparrow}^{\dag},c_{0\uparrow})T_{\text{eff}}\begin{pmatrix}
    f_{\uparrow} \\
    f_{\downarrow} \\
    f_{\uparrow}^{\dag}\\
    f_{\downarrow}^{\dag}
\end{pmatrix},
\end{equation}
where
\begin{equation}
    T_{\text{eff}}=it\begin{pmatrix}
    1 & 0 & \cos \Theta e^{i\Phi} & -i\sin \Theta\\
    \cos \Theta e^{-i\Phi} & i\sin \Theta& 1 & 0
\end{pmatrix}.
\end{equation}
This coupling term would induce the self-energy:
\begin{equation}
    \Sigma_R^r=\frac{2t^2}{E^+}(1+\cos\Theta\cos\Phi\tau_x-\cos\Theta\sin\Phi\tau_y).
\end{equation}
At the limit of $E\to0$, the retarded Green function can be deduced similarly to Eq. (\ref{eq8}) as:
\begin{equation}
    G^r(E\to0)\approx
     \frac{-1+\cos\Theta\cos\Phi\tau_x-\cos\Theta\sin\Phi\tau_y}{-2i\sin k_F+\frac{2t^2}{E^{+}}\sin^2\Theta}.
\end{equation}
When $\Theta=0$, which means $\bm{d}$ is within the $xy$-plane, the $E$ relevant term vanishes, and the Andreev and normal reflection amplitudes can be deduced as:
\begin{equation}
    r_{he}(E=0)=-e^{-i\Phi},\ 
    r_{ee}(E=0)=0.
\end{equation}
When $\Theta\ne0$, the retarded Green function is a null matrix when $E=0$, so the reflection amplitudes can be easily acquired: $r_{he}(E=0)=0$, $r_{ee}(E=0)=-1$.

To summarize, at $E=0$, when the spin-polarized chain is coupled to a time-reversal symmetric $p$-wave superconductor, the $\bm{d}$-vector orientation dependences of the reflection amplitudes are:
\begin{equation}
    \begin{aligned}
    \begin{cases}
         r^{\uparrow\uparrow}_{he}(E=0)=-e^{-i\Phi}, \ \ r^{\uparrow\uparrow}_{ee}(E=0)=0,&\Theta=0;\\
         r^{\uparrow\uparrow}_{he}(E=0)=0,\ \ \ \ \ r^{\uparrow\uparrow}_{ee}(E=0)=-1, &\Theta\ne0.
    \end{cases}
    \end{aligned}
\end{equation}

When $E\rightarrow 0$, the retarded Green functions take the following expressions:
\begin{equation}
    \begin{aligned}
        &G_{he}^r(E)=\frac{2t^2/E }{-e^{i(k_h-k_e)}-2(e^{ik_h}-e^{-ik_e})t^2/E+\frac{4t^4}{E^2}\sin^2\Theta}\cos \Theta e^{-i\Phi},\\
        &G_{ee}^r(E)=\frac{e^{ik_h}-2t^2/E }{-e^{i(k_h-k_e)}-2(e^{ik_h}-e^{-ik_e})t^2/E+\frac{4t^4}{E^2}\sin^2\Theta}.
    \end{aligned}
\end{equation}
When $\Theta=0$, these retarded Green functions are analogous to those in Eq. (\ref{eq29}) and we just need to replace $\Phi$ by $-\Phi$, and $t'$ by $t$. The reflection amplitudes and coefficients can be expanded as ($t\ll1$): 
\begin{equation}
    \begin{aligned}
        &r_{he}(E)\approx(-1-i\frac{E}{2v_Ft^2}+\frac{E^2}{4v_F^2t^4})e^{-i\Phi},\\
        &r_{ee}(E)\approx i\frac{e^{2ik_F}}{2v_Ft^2}E,\\
        &R_A(E) \approx1-(\frac{E}{2v_Ft^2})^2,\\
    &R_N(E) = 1-R_A(E).
    \end{aligned}
\end{equation}
When $\Theta\ne0$ the Andreev and normal reflection amplitudes and coefficients take the following approximations:
\begin{equation}
\begin{aligned}
    &r_{he}(E)\approx i\frac{v_F\cos \Theta}{2t^2\sin^2 \Theta}Ee^{-i\Phi},\\
    &r_{ee}(E)\approx -1-i\frac{v_F}{2t^2\sin^2 \Theta}E+\frac{iv_Fe^{ik_F}\sin^2\Theta+v_F^2}{4t^4\sin^4\Theta}E^2,\\
    &R_A(E)\approx (\frac{v_F\cos\Theta E}{2t^2\sin^2\Theta})^2,\\
    &R_N(E)= 1-R_A(E).
\end{aligned} 
\end{equation}
If $\Theta\to 0$, the reflection amplitudes and coefficients can be further simplified as:
\begin{equation}
    \begin{aligned}
        &r_{he}(E)\approx i\frac{v_F}{2t^2\Theta^2}Ee^{-i\Phi},\\
        &r_{ee}(E)\approx -1-i\frac{v_F}{2t^2\Theta^2}E+\frac{v_F^2}{4t^4\Theta^4}E^2,\\
        &R_A(E)\approx (\frac{v_FE}{2t^2\Theta^2})^2.
    \end{aligned}
\end{equation}

In this case, the $E$-dependent charge and spin pumping can be obtained as ($t\ll1$):
\begin{equation}
    \begin{cases}
        \mathcal{Q}(E)/e=\mathcal{S}_z(E)/\frac{\hbar}{2}\approx 1-(\frac{E}{2v_Ft^2})^2, &\alpha=0;\\
        \mathcal{Q}(E)/e=\mathcal{S}_z(E)/\frac{\hbar}{2}\approx(\frac{v_F\cos\alpha E}{2t^2\sin^2\alpha})^2\xrightarrow{\alpha\to0}(\frac{v_FE}{2t^2\alpha^2})^2, &\alpha\ne0.      
    \end{cases}
\end{equation}

\section*{IV. The influence of the interference effect on the transport}

\begin{figure}[ht]
  \begin{center}
	\includegraphics[width=12cm,height=3.2cm]{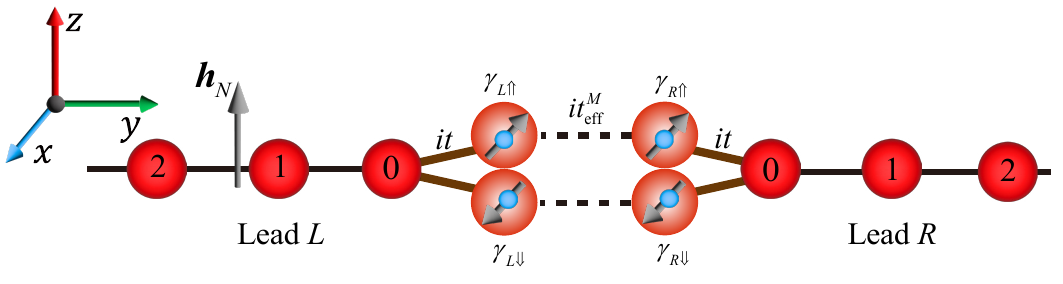}	
  \end{center}
  \vspace{-0.4cm}
	\caption{ Effective model of a time-reversal symmetric $p$-wave superconductor with finite length, coupled at two ends with two spinful leads, labeled by Lead $L$ and Lead $R$, respectively. Each normal lead is coupled at the interface to two bounded Kramers degenerate Majorana fermions. Here $it$ is the coupling constant between the lead and the Majorana fermions, $it^M_{\text{eff}}$ is the effective coupling integral between the Majorana fermions at two ends, and $\bm{h}_N$ is the Zeeman field applied in the left normal lead.
  } \label{figa3}
\end{figure}

When 1D $p$-wave superconductor has a relatively small length $L$, the interference effect between the Majorana fermions at both ends will observably influence the pumping properties for the system discussed in the main text.
Here we explain the results of the pumping by viewing the superconductor as two Kramers pairs of Majorana fermions at both ends, coupled with each other by an effective coupling strength $it^M_{\text{eff}}=it^Me^{-L/l_M}$, with $l_M$ the evanescent length of the Majorana zero modes. As before, the coupling strength between Majorana fermions and the 1D leads is still $it$. At left end of the $p$-wave superconductor, the Kramers pair of Majorana fermions can be expressed as $\gamma_{L\Uparrow}=e^{-i\alpha/2}f_{L\Uparrow}+e^{i\alpha/2}f_{L\Uparrow}^{\dag}$ and $\gamma_{L\Downarrow}=e^{i\alpha/2}f_{L\Downarrow}+e^{-i\alpha/2}f_{L\Downarrow}$, where $\alpha$ is relevant to the orientation of the $p$-wave $\boldsymbol{d}$-vector. Due to $p$-wave spin-triplet pairing symmetry, Majorana pairs at two ends acquire additional $\pi/2$ phase difference. So we have at the right end: $\gamma_{R\Uparrow}=-ie^{-i\alpha/2}f_{R\Uparrow}+ie^{i\alpha/2}f_{R\Uparrow}^{\dag}$ and $\gamma_{R\Downarrow}=-ie^{i\alpha/2}f_{R\Downarrow}+ie^{-i\alpha/2}f_{R\Downarrow}$. Thus the coupling term at the left interface is:
\begin{equation}
\begin{aligned}
    H_T^L&=it[(e^{-i\alpha/2}c_{0\Uparrow}+e^{i\alpha/2}c_{0\Uparrow}^{\dag})\gamma_{L\Uparrow}+(e^{i\alpha/2}c_{0\Downarrow}+e^{-i\alpha/2}c_{0\Downarrow}^{\dag})\gamma_{L\Downarrow}]\\   &=(c_{0\Uparrow}^{\dag},c_{0\Downarrow}^{\dag},c_{0\Uparrow},c_{0\Downarrow})\ itT_{\text{eff}}\ \begin{pmatrix}
    f_{L\Uparrow} \\
    f_{L\Downarrow} \\
    f_{L\Uparrow}^{\dag}\\
    f_{L\Downarrow}^{\dag}
\end{pmatrix},
\end{aligned}
\end{equation}
where
\begin{equation}
T_{\text{eff}}=\begin{pmatrix}
    1 & 0 & e^{i\alpha} & 0\\
    0 & 1 & 0 & e^{-i\alpha}\\
    e^{-i\alpha} & 0 & 1 & 0\\
    0 & e^{i\alpha} & 0 & 1
\end{pmatrix}.  
\end{equation}
Since the $\pi/2$ phase can be absorbed by fermion operators $f_{R\Uparrow}$ and $f_{R\Downarrow}$, the coupling term at the right interface $H_T^R$ shares exactly the same form with the same coulping matrix $T_{\text{eff}}$. The coupling term between the Majorana fermions at two ends is assumed to be:
\begin{equation}
\begin{aligned}
    H_T^M&=it_{\text{eff}}^M(\gamma_{L\Uparrow}\gamma_{R\Uparrow}+\gamma_{L\Downarrow}\gamma_{R\Downarrow})\\
    &=(f_{L\Uparrow}^{\dag},f_{L\Downarrow}^{\dag},f_{L\Uparrow},f_{L\Downarrow})\ it_{\text{eff}}^MT_{\text{eff}}\ \begin{pmatrix}
    f_{R\Uparrow} \\
    f_{R\Downarrow} \\
    f_{R\Uparrow}^{\dag}\\
    f_{R\Downarrow}^{\dag}
\end{pmatrix}
\end{aligned}.
\end{equation}
\iffalse
where
\begin{equation}
T_{\text{eff}}^\text{M}=t_{\text{eff}}^\text{M}\begin{pmatrix}
    -i & 0 & ie^{i\alpha} & 0\\
    0 & -i & 0 & ie^{-i\alpha}\\
    -ie^{-i\alpha} & 0 & i & 0\\
    0 & -ie^{i\alpha} & 0 & i
\end{pmatrix}.
\end{equation}
\fi
No coupling between Majorana fermions with different spins is assumed because $\Uparrow\Uparrow$ and $\Downarrow\Downarrow$ pairing channels are decoupled in a time-reversal symmetric $p$-wave superconductor.

We now demonstrate that the main physics of the quantum transport in a realistic system including the differential conductance in a stationary case or the pumping properties in a periodically driven case can be captured by the effective model. Consider the scattering processes that at fixed energy $E$, a Lead-$L$ incident electron is reflected, or a Lead-$R$ incident electron is transmitted to the left chain. Based on the effective model, the scattering amplitudes can be acquired via the extended Fisher-Lee relation in Eq. (\ref{eq9}):
\begin{equation}
    S_{\alpha\beta}^{pq}(E)=-\delta_{pq}\delta_{\alpha\beta}+i[\Gamma_{\alpha\alpha}^{p}(E)]^{1/2}G^{pq}_{\alpha\beta}(E)[\Gamma_{\beta\beta}^{q}(E)]^{1/2}.
\end{equation}
Here $S_{\alpha\beta}^{pq}(E)$ represents the scattering amplitude where the incident particle with state $\beta$ in lead $q$ is scattered as $\alpha$ in lead $p$ with $\alpha,\beta\in$ $\{e$$\Uparrow$, $e$$\Downarrow$, $h$$\Uparrow$, $h$$\Downarrow\}$ and $p,q \in \{L,R\}$. The retarded Green function is still given by $G^r(E)=(E^+-\bm{H}-\Sigma_L^r-\Sigma_R^r)^{-1}$, where $\bm{H}$ now stands for the matrix Hamiltonian of the central scattering region including the superconductor and two normal sites $0$ in both chains:
\begin{equation}
    \bm{H}=\begin{pmatrix}
    -\mu\tau_z & itT_{\text{eff}} & 0 & 0\\
    -itT_{\text{eff}} & 0 & it_{\text{eff}}^MT_{\text{eff}} & 0\\
    0 & -it_{\text{eff}}^MT_{\text{eff}} & 0 & -itT_{\text{eff}}\\
    0 & 0 & itT_{\text{eff}} & -\mu\tau_z
\end{pmatrix}.
\end{equation}

\begin{figure}[!bt]
  \begin{center}
	\includegraphics[width=18cm,height=7.85cm]{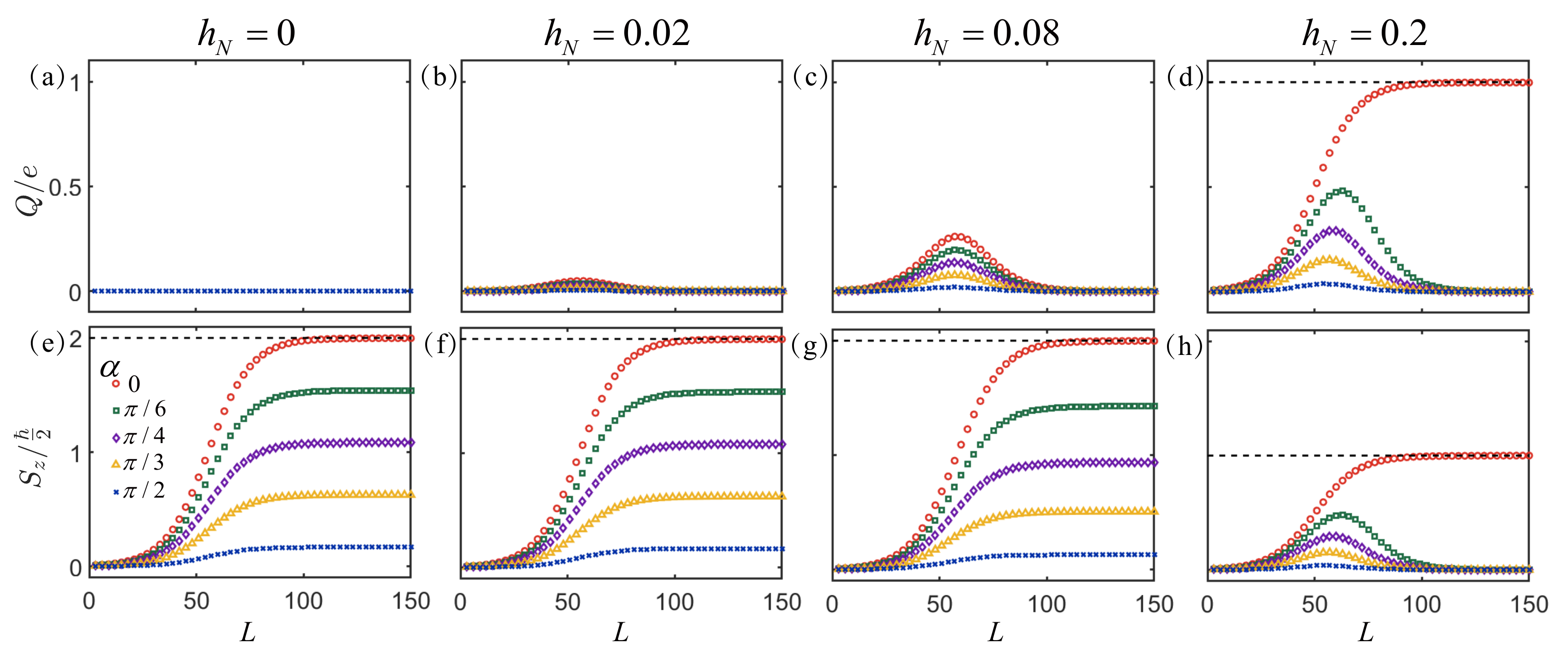}	
  \end{center}
  \vspace{-0.4cm}
	\caption{(a)-(d) Periodically pumped charge and (e)-(h) corresponding pumped spin in one cycle in the left lead as functions of length $L$ for different Zeeman field $h_N$ with $l_M=20$ and $t^M=t=0.1$. Lead $L$ is fully spin-polarized when $h_N>0.1$.
  } \label{figa4}
\end{figure}

In the periodically driving process, the spin-polarization of the $p$-wave superconductor is modulated adiabatically by the varying tiny magnetic field $\bm{B}_S(t)$ and instantly aligns with it, as mentioned in the main text. The magnetic field $\boldsymbol{B}_S(t)$ at each moment takes the orientation $(\theta_B(t),\phi_B(t))$. By performing the spin rotation: $(c_{\Uparrow}^{\dag},c_{\Downarrow}^{\dag})=(c_{\uparrow}^{\dag},c_{\downarrow}^{\dag})U$, with $U(t)=U_z(\phi_B(t))U_y(\theta_B(t))$, the effective hopping matrix becomes:
\begin{equation}
    T_{\text{eff}}\longrightarrow\begin{pmatrix}
    U(t) & 0 \\
    0 & U^*(t)
\end{pmatrix}T_{\text{eff}}\begin{pmatrix}
    U^{\dag}(t) & 0 \\
    0 & U^T(t)
\end{pmatrix}.
\end{equation}
Now the scattering amplitudes can be calculated at each moment and we can deduce the charge and spin pumping using the equation introduced in the main text. In Figs. \ref{figa4}(a)-(d) we show the pumped charge in one cycle in the left lead as functions of length $L$ for different $\boldsymbol{d}$-vector orientations characterized by $\alpha$. When $h_N=0$, no pumped charge is found and as $h_N$ increases the pumped charge becomes finite and forms a peak at a certain length $L$ proportional to $l_M$. When the left chain is fully polarized, the pumped charge for $\alpha=0$ becomes quantized at $Q=e$ for sufficient large $L$. Figs. \ref{figa4}(e)-(h) show the corresponding pumped spin. When the left chain is partially spin-polarized, the pumped spin starts from zero at small $L$ and at large enough $L$ is approaching a fixed value which sensitively depends on $\alpha$, agreeing well with the results of the realistic system discussed in the main text. When the left chain is fully spin-polarized, the quantized pumped spin at $\alpha=0$ abruptly changes from $S_z=\hbar$ to $S_z=\hbar/2$, which also agrees well with those in the main text. Notice that periodic oscillation with period $\pi/k_F$ is not captured in this effective model. 
\end{document}